\documentclass[sigplan,nonacm]{acmart}
\usepackage{siunitx}
\usepackage{courier}
\usepackage{comment}
\usepackage{helvet}
\usepackage{xspace}
\usepackage{booktabs}
\usepackage{algorithm}
\usepackage{algorithmic}
\usepackage{subcaption}

\newcommand{\sys}{CXLAimPod\xspace}

\renewcommand\footnotetextcopyrightpermission[1]{}
\begin{abstract}

The proliferation of data-intensive applications, ranging from large language models to key-value stores, increasingly stresses memory systems with mixed read-write access patterns. Traditional half-duplex architectures such as DDR5 are ill-suited for such workloads, suffering bus turnaround penalties that reduce their effective bandwidth under mixed read-write patterns. Compute Express Link (CXL) offers a breakthrough with its full-duplex channels, yet this architectural potential remains untapped as existing software stacks are oblivious to this capability.

This paper introduces \sys, an adaptive scheduling framework designed to bridge this software-hardware gap through system support including cgroup-based hints for application-aware optimization. Our characterization quantifies the opportunity, revealing that CXL systems achieve 55-61\% bandwidth improvement at balanced read-write ratios compared to flat DDR5 performance, demonstrating the benefits of full-duplex architecture. To realize this potential, the \sys framework integrates multiple scheduling strategies with a cgroup-based hint mechanism to navigate the trade-offs between throughput, latency, and overhead. Implemented efficiently within the Linux kernel via eBPF, \sys delivers significant performance improvements over default CXL configurations. Evaluation on diverse workloads shows 7.4\% average improvement for Redis (with up to 150\% for specific sequential patterns), 71.6\% improvement for LLM text generation, and 9.1\% for vector databases, demonstrating that duplex-aware scheduling can effectively exploit CXL's architectural advantages.
\end{abstract}

\begin{document}

\title{\sys: CXL Memory is all you need in AI era}

% \author{Anonymous Authors}
\author{Yiwei Yang$^{=}$, Yusheng Zheng$^{=}$, Yiqi Chen, Zheng Liang, Kexin Chu, Zhe Zhou, Andi Quinn, Wei Zhang}

\maketitle

\section{Introduction}
\label{sec:intro}

Modern applications are becoming increasingly data-hungry. Large language models need hundreds of gigabytes for their parameters~\cite{brown2020language}, while distributed databases manage petabytes of data~\cite{decandia2007dynamo}. These applications constantly mix reads and writes to memory, making the "memory wall" (the speed gap between processors and memory) worse than ever~\cite{wulf1995hitting}. This is a significant problem: studies have shown that recommendation systems can spend over half their time waiting for memory~\cite{eisenman2018bandana}, and search systems can waste substantial processing cycles on memory delays~\cite{kanev2015profiling}. These delays cost money: memory systems now consume up to 40\% of datacenter power.

The root cause is how traditional memory works. For decades, DRAM has used a shared bus that can only handle reads or writes at one time, not both~\cite{jacob2018memory}. Even though DDR5 can theoretically deliver 76.8 GB/s per channel~\cite{jedec2020ddr5}, it struggles with today's applications that constantly switch between reading and writing data. The problem is the delay when switching directions on the bus~\cite{hassan2016chargecache}. Each switch takes 15-20 cycles (about 11-15 nanoseconds), which seems small but adds up quickly. Measurements show DDR5 performance remains relatively flat across different read-write ratios, varying by only 26\% (153-189 GB/s) due to bus turnaround penalties. To work around this, engineers have created complex workarounds: memory controllers batch similar operations together, and programmers rewrite their code to separate reads from writes, making software more complex for marginal gains.

The Compute Express Link (CXL) standard offers a potential solution through a different architectural paradigm~\cite{sharma2023introduction}. Beyond its benefits for coherent memory expansion and pooling~\cite{pond2023pond}, CXL builds on the PCIe physical layer to provide physically separate, full-duplex channels that enable concurrent bidirectional data transfers~\cite{pcie2022specification}. This architectural shift eliminates the half-duplex bottleneck that has limited memory system efficiency for decades. Characterization shows that while DDR5 bandwidth remains relatively flat across read-write ratios, CXL achieves 55-61\% bandwidth improvement at balanced read-write ratios, demonstrating significant duplex benefits. However, this potential remains largely untapped. Existing operating systems and their memory schedulers, designed for a half-duplex world, are unaware of CXL's duplex capabilities. Without intelligent software management, CXL devices cannot fully exploit their architectural advantages.

To bridge this software-hardware gap, we present \sys, an adaptive scheduling framework that helps exploit the performance potential of CXL memory. \sys intercepts memory requests and intelligently co-schedules them to exploit CXL's full-duplex link for simultaneous bidirectional transfers, maximizing concurrent utilization while minimizing interference~\cite{mutlu2008parallelism}. The framework is designed for production deployment, overcoming the challenges of kernel modification through the use of eBPF for safe, efficient in-kernel extension~\cite{gregg2019bpf}.

An important feature of \sys is the integration of cgroup-based hints that enable application-level guidance of allocation and scheduling decisions~\cite{menage2007cgroups}. This mechanism allows system administrators and applications to communicate their memory access patterns and performance requirements through the familiar cgroup interface, enabling the scheduler to make informed decisions without requiring application modification. The framework incorporates multiple scheduling strategies that explore different trade-offs between performance and overhead, from simple threshold-based approaches to more sophisticated adaptive algorithms~\cite{astrom2010feedback,hur2007adaptive,hashemi2018learning}, along with cgroup-aware scheduling for multi-tenant optimization~\cite{verma2015large}.

This paper makes the following key contributions:
\begin{itemize}
\item We provide an in-depth characterization of CXL's full-duplex performance under mixed read-write workloads.
\item We design and implement \sys, a duplex-aware scheduling framework for Linux that is both high-performance and safely deployable without kernel modifications using eBPF.
\item We demonstrate through evaluation with microbenchmarks and real-world applications, including Redis, RocksDB, graph analytics, and a 671B-parameter language model, that \sys delivers 7.4\% average improvement for Redis (up to 150\% for sequential patterns), 71.6\% improvement for LLM text generation, and 9.1\% for vector databases over default CXL configurations.
\end{itemize}

The remainder of this paper is organized as follows. Section \ref{sec:background} provides background on CXL and memory scheduling. Section \ref{sec:characterization} presents our performance characterization. Section \ref{sec:design} describes the design of our scheduling framework. Section \ref{sec:implementation} details the \sys implementation. Section \ref{sec:evaluation} presents our evaluation, and we conclude in Section \ref{sec:conclusion}.
\section{Background and Motivation}

This section examines the limitations of DDR architectures and the opportunities presented by CXL's full-duplex design, providing context for why existing software fails to exploit modern hardware capabilities.

\subsection{The Half-Duplex Legacy of DDR Memory}

DDR5 achieves 76.8 GB/s per channel through high signaling rates and wide data paths~\cite{jedec2020ddr5}. However, it retains the half-duplex, shared-bus architecture from earlier DRAM interfaces~\cite{jacob2018memory}. The shared command and data buses require explicit coordination when switching between reads and writes~\cite{hassan2016chargecache}. This bus turnaround penalty (15-20 cycles on DDR5 systems) creates idle time where no useful work occurs. At 6400 MT/s, this translates to 11.25-15 nanoseconds of idle time per direction switch.

\subsection{CXL: A Full-Duplex Architecture}

In contrast to DDR's limitations, CXL introduces a fundamentally different approach. It builds on PCIe 5.0's physical layer with a protocol stack optimized for memory~\cite{sharma2023introduction,cxl2022specification30}. Each PCIe lane operates at 32 GT/s with 128b/130b encoding, yielding 31.5 Gb/s per direction with under 2\% overhead~\cite{pcie2022specification}. The key distinction lies in CXL's use of separate transmit (TX) and receive (RX) paths inherited from PCIe's differential signaling~\cite{gouk2022direct}. Each lane has dedicated differential pairs for TX and RX with no electrical connection between them, maintaining independence through SerDes blocks, clock domains, and protocol processing.

This architecture enables the x16 configuration to provide up to 64 GB/s in each direction through 16 lanes at 4 GB/s each, totaling 128 GB/s bidirectional bandwidth capability for simultaneous transfers, though actual device implementations may introduce bottlenecks~\cite{sun2023demystifying}. CXL 2.0 implements three protocols: CXL.io for PCIe compatibility and device discovery, CXL.cache for device-initiated coherent requests, and CXL.mem for host memory requests to device memory~\cite{pond2023pond}.

By reducing turnaround penalties through physical channel separation, CXL achieves improved bandwidth efficiency~\cite{sun2023demystifying}. While CXL latency (130-200ns) exceeds DDR's (75-85ns), the bandwidth efficiency gains justify this trade-off for bandwidth-sensitive datacenter workloads~\cite{gouk2022direct}.

\subsection{The Software-Hardware Gap}

Despite CXL's hardware advances, an important challenge remains: system software still assumes half-duplex operation~\cite{gu2024vortex}, creating a gap that limits applications from exploiting these capabilities. Memory controllers minimize direction switches rather than maximize concurrent utilization~\cite{mutlu2008parallelism}, using batching to group operations. While this mitigates DDR turnaround penalties, it prevents CXL's concurrent operation. Similarly, OS allocators focus on NUMA locality without considering duplex capabilities~\cite{dashti2013traffic}. Applications further compound the problem by separating read and write phases, assuming resource competition.

Addressing this gap requires careful consideration of overhead. Instrumenting every memory operation would be prohibitive~\cite{hennessy2019new}. Instead, PMU-based profiling with intelligent scheduling offers a practical solution~\cite{ghose2019workload}. This approach captures access patterns without full instrumentation, enabling informed decisions while maintaining compatibility.

\subsection{Memory Access Patterns in Modern Applications}

To understand why duplex-aware scheduling matters, we must examine how modern applications access memory. Datacenter workloads exhibit three distinct patterns that interact differently with memory architectures. First, read-dominated workloads like analytics, ML inference, and graph algorithms generate 80-95\% reads~\cite{eisenman2018bandana}, processing datasets beyond cache capacity. Even these workloads face disruption from occasional writes for stack operations, spills, and metadata. In DDR systems, these writes interrupt read streams, destroying row buffer locality and prefetch efficiency.

Second, write-intensive workloads including logging, stream processing, and write-ahead logs generate 70-90\% writes~\cite{ongaro2014search}. In DDR systems, write buffers fill quickly, causing pipeline stalls and latency variations. CXL's full-duplex link enables simultaneous bidirectional transfers, potentially improving write throughput when reads are also present.

Third, mixed workloads from key-value stores and databases with balanced read-write ratios present challenges for DDR but opportunities for CXL~\cite{cooper2010benchmarking}. Redis exemplifies this challenge, achieving only 45\% DDR5 bandwidth utilization due to constant switching~\cite{carlson2013redis}. Fine-grained operations (64B-1KB) make turnaround penalties nearly equal data transfer time.

These patterns often coexist within single applications. Web platforms combine read-heavy browsing, write-intensive orders, and mixed inventory updates. Where DDR forces over-provisioning or performance loss, CXL's full-duplex design transforms mixed workloads into efficiency opportunities. This potential motivates the development of duplex-aware scheduling frameworks that can dynamically adapt to changing access patterns.

\section{Duplex Characterization}
\label{sec:characterization}

\begin{figure}[htbp]
    \centering
\includegraphics[width=1.05\columnwidth]{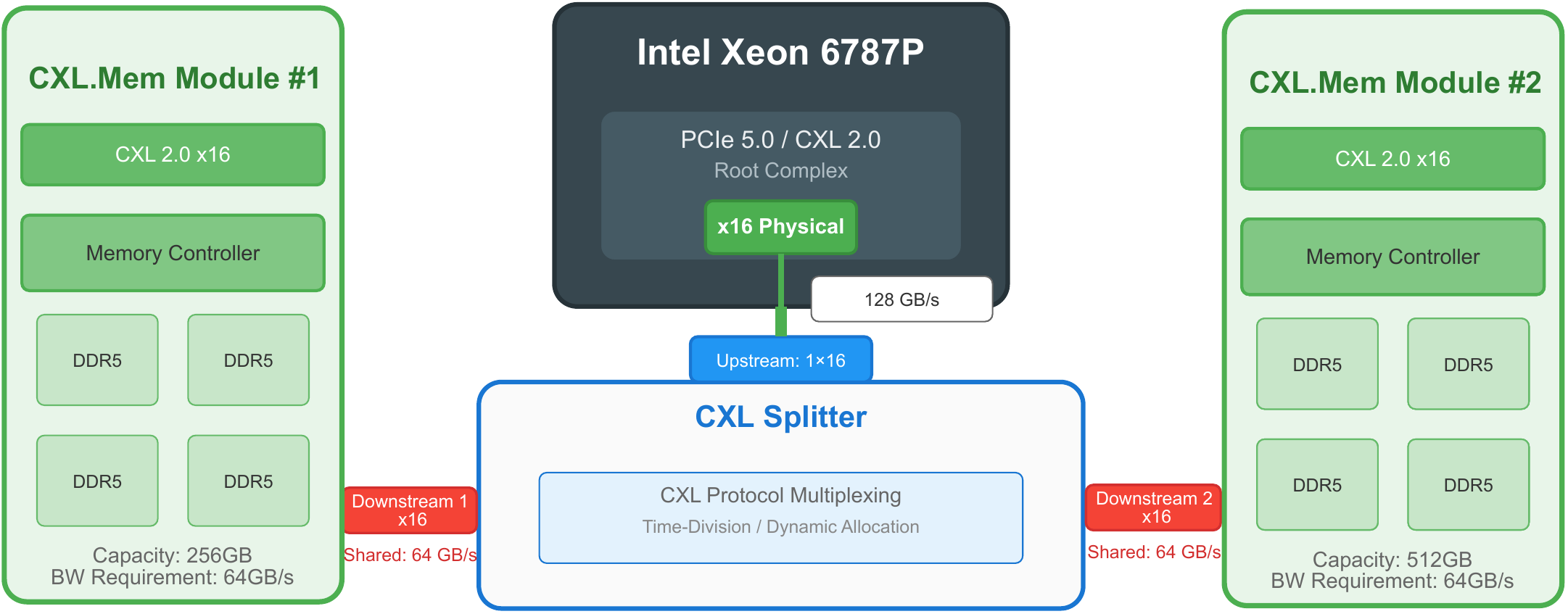}
    \caption{\sys Architecture}
    \label{fig:cxl_vs_ddr}
\end{figure}

 % \begin{figure}
 %       \includegraphics[width=0.8\columnwidth]{img/sys.pdf}
 %       \label{fig:scale}
 %       \caption{\sys Real Device}
 %   \end{figure}

We conduct characterization of CXL's duplex behavior to understand its performance characteristics under diverse conditions and identify optimization opportunities. The experimental methodology employs controlled measurements to isolate specific aspects of duplex behavior while accounting for variables that affect memory system performance.

\subsection{Experimental Methodology}

The experimental platform uses a single-socket Intel Xeon 6 6787P (Granite Rapids) with 86 cores at 2.0 GHz base frequency and 3.8 GHz turbo. The cache hierarchy includes 48 KiB L1d and 64 KiB L1i per core, 2 MiB L2 per core (172 MiB total), and 336 MiB shared L3. The system has 128GB DDR5 memory across two NUMA nodes (64GB each) as our baseline.

For CXL, we use two Montage Technology memory expanders (256GB and 512GB) with DDR5 backing storage and custom controller ASICs, plus two Astera Labs switches/retimers for signal integrity. These connect via PCIe Gen5 x16 with CXL 2.0 protocol support. The CXL memory appears as memory-only NUMA nodes without CPU affinity, which is standard for Type 3 expanders.

We developed a custom microbenchmark to evaluate duplex behavior with precise control over workload characteristics. The benchmark uses multi-threaded workers performing sequential or random operations on NUMA-pinned memory. Key parameters include read-write ratios (0-100\% reads), block sizes (4KB-1MB), bandwidth limits via token-bucket rate limiting, and thread counts (up to 500). Each worker tracks operation statistics while a coordinator computes aggregate metrics (bandwidth, IOPS, efficiency). This isolates specific workload impacts on duplex performance while maintaining realistic access patterns.

Table~\ref{tab:numa_cxl} shows the NUMA topology with nodes 0-1 hosting DDR5 with CPUs and nodes 2-3 as memory-only CXL. NUMA distances indicate relative latencies: local access (baseline), cross-socket DDR5 (20\% higher), CXL from CPU nodes (40\% higher), and inter-CXL (60\% higher), reflecting CXL's protocol overhead.

For monitoring, we also use Intel uncore PMUs for cycle-level memory controller events and CXL device counters via memory-mapped registers for channel utilization and protocol efficiency metrics. The 896GB total memory (128GB DDR5 + 768GB CXL) enables testing workloads beyond DDR5 capacity limits.

\begin{figure*}[t]
  \centering

  \begin{subfigure}{\textwidth}
    \centering
    \includegraphics[width=\linewidth]{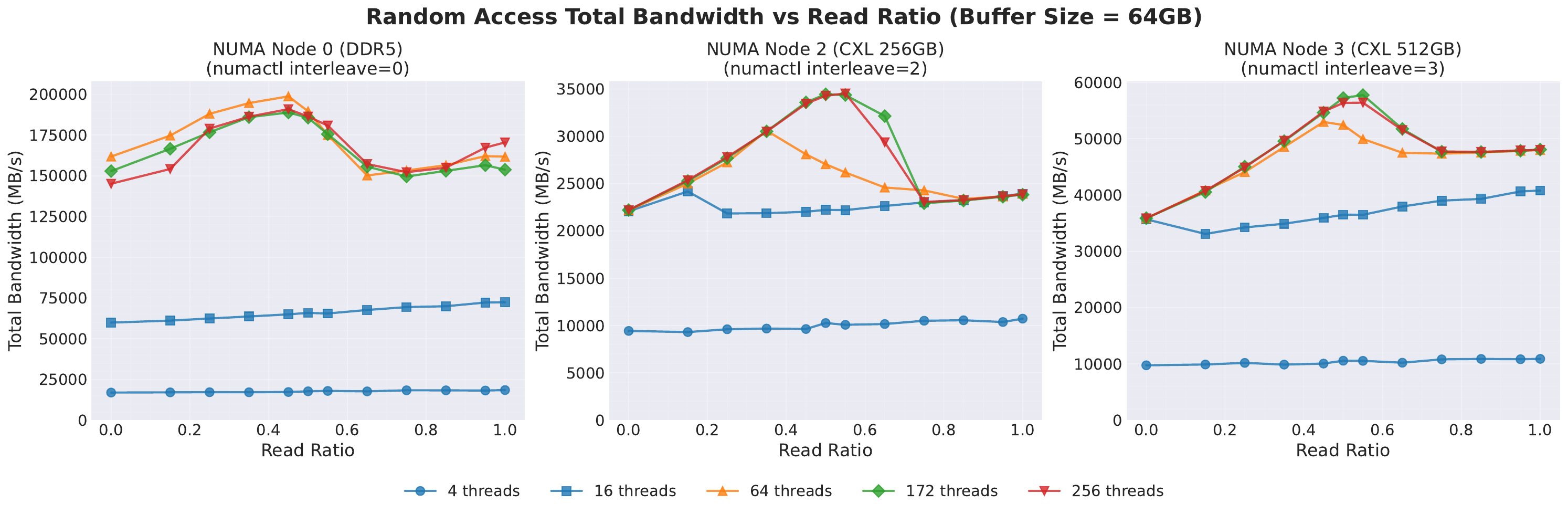}
    \caption{CXL Duplex Memory Showcase}
    \label{fig:showcase}
  \end{subfigure}

  \bigskip % or \vspace{1em}

  \begin{subfigure}{\textwidth}
    \centering
    \includegraphics[width=\linewidth]{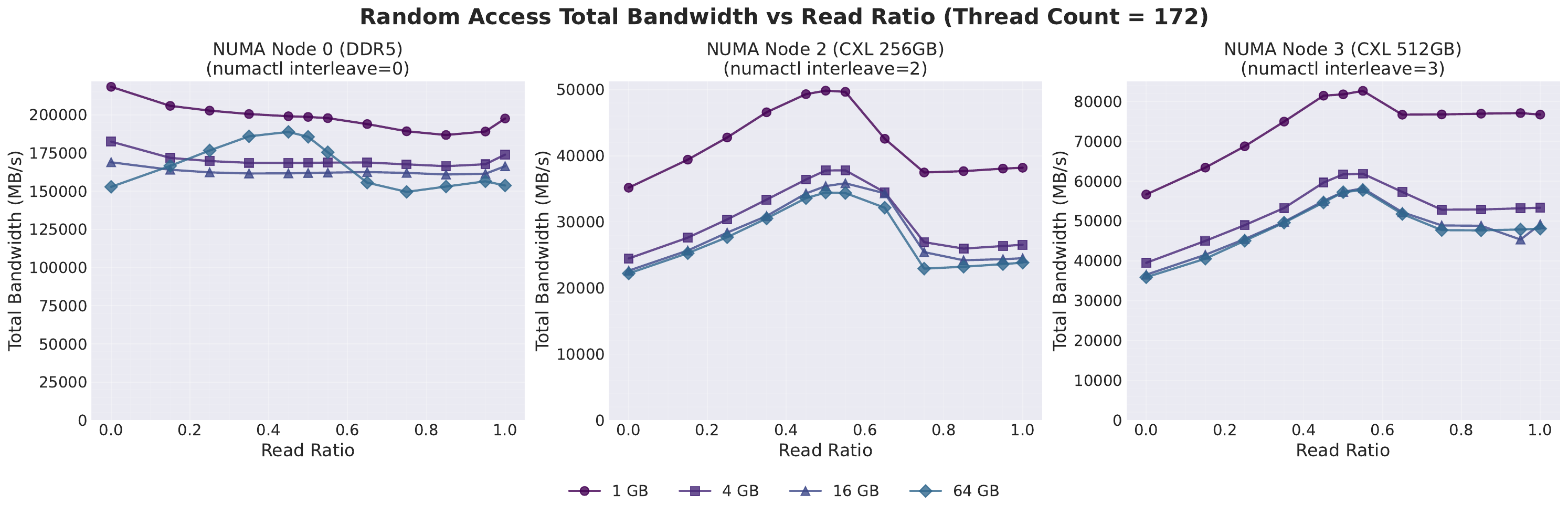}
    \caption{CXL Scalability and IOPS}
    \label{fig:scale}
  \end{subfigure}

  \caption{CXL System Performance Characteristics}
  \label{fig:combined}
\end{figure*}

\begin{table}[h!]
\centering
\caption{NUMA and CXL Memory Configuration}
\label{tab:numa_cxl}
\begin{tabular}{|c|c|c|c|c|}
\hline
\textbf{Node} & \textbf{Cores} & \textbf{MemSize} & \textbf{MemType} & \textbf{Status} \\ \hline
Node 0 & 0--42 & 64 GB  & DDR5 (Local) & with CPUs \\ \hline
Node 1 & 43--85 & 64 GB  & DDR5 (Local) & with CPUs \\ \hline
Node 2 & None              & 256 GB & CXL Memory   & Mem-only \\ \hline
Node 3 & None              & 512 GB & CXL Memory   & Mem-only \\ \hline
\end{tabular}
\end{table}

% \begin{table}[h!]
% \centering
% \caption{NUMA Distance Matrix (Relative Latency)}
% \label{tab:numa_distance}
% \begin{tabular}{|c|c|c|c|c|}
% \hline
%        & \textbf{Node 0} & \textbf{Node 1} & \textbf{Node 2} & \textbf{Node 3} \\ \hline
% \textbf{Node 0} & 10 & 12 & 14 & 14 \\ \hline
% \textbf{Node 1} & 12 & 10 & 14 & 14 \\ \hline
% \textbf{Node 2} & 14 & 14 & 10 & 16 \\ \hline
% \textbf{Node 3} & 14 & 14 & 16 & 10 \\ \hline
% \end{tabular}
% \end{table}

\subsection{Random Access Bandwidth Analysis}

The characterization reveals key observations about CXL's duplex behavior compared to DDR5 under varying read-write workloads with random access patterns. We evaluate memory systems using 4-172 concurrent threads to ensure sufficient load generation, testing both small (1GB) and large (64GB) buffer sizes to capture different caching behaviors. Random access patterns stress the memory system by eliminating spatial locality benefits and maximizing DRAM row buffer misses.

\textbf{Observation 0: CXL achieves 17-29\% of DDR5's bandwidth due to higher latency and protocol overhead.} With 64GB buffers under random access, DDR5 achieves an average of 166.7 GB/s while the 256GB CXL device manages only 27.8 GB/s (17\% of DDR5) and the 512GB device reaches 48.6 GB/s (29\% of DDR5). This 3.4-6× bandwidth gap stems from CXL's higher latency (130-200ns vs 75-85ns), PCIe protocol overhead, and the additional indirection through the CXL controller. Despite lower absolute bandwidth, CXL provides value through memory expansion beyond DDR5 capacity limits and better bandwidth efficiency under mixed workloads, making it suitable for capacity-sensitive rather than bandwidth-critical applications.

\textbf{Observation 1: CXL achieves 55-61\% bandwidth improvement at balanced read-write ratios while DDR5 remains flat.} For random access with 64GB buffers, DDR5 bandwidth varies by only 26\% across all read-write ratios (153-189 GB/s), limited by bus turnaround penalties. In contrast, CXL devices show pronounced duplex benefits: the 256GB device achieves 34.4 GB/s at 50\% read ratio versus 22.2 GB/s for pure writes (55\% improvement), while the 512GB device reaches 57.8 GB/s at 55\% read ratio versus 35.9 GB/s for pure writes (61\% improvement). This validates that CXL's full-duplex link provides architectural advantages for mixed workloads that DDR5's half-duplex design cannot match.

\textbf{Observation 2: Write performance asymmetry differs between CXL and DDR5.} CXL devices show consistent write performance penalties, with writes achieving 74-93\% of read throughput due to protocol overhead and coherency requirements. The 512GB CXL device exhibits the largest gap (0.75× write/read ratio), while the 256GB device maintains 0.93×. DDR5, however, shows minimal read-write performance difference with random access, achieving near parity (0.99× ratio) due to its optimized memory controller that batches operations to minimize turnaround overhead. This asymmetry has implications for scheduling: CXL benefits from read-prioritization while DDR5 can treat both operations equally.

\textbf{Observation 3: Buffer size amplifies CXL bandwidth but preserves duplex characteristics.} Reducing buffer size from 64GB to 1GB increases bandwidth across all memory types: DDR5 improves by 19\% (167 to 198 GB/s), while CXL devices show 52-53\% gains. The 512GB CXL device jumps from 48.6 to 74.5 GB/s with smaller buffers, approaching DDR5's performance for cache-friendly workloads. Despite these absolute changes, the relative duplex behavior remains consistent: CXL still peaks at 50-55\% read ratio regardless of buffer size, confirming that duplex benefits are architectural rather than workload-dependent.

\textbf{Observation 4: CXL bandwidth saturates at 172 threads while DDR5 peaks at 64 threads.} Thread scaling analysis reveals distinct saturation points: DDR5 reaches peak bandwidth of 198.8 GB/s with 64 threads and shows minimal improvement beyond (less than 5\% gain from 64 to 256 threads), while CXL devices require 172 threads to achieve maximum throughput (34.4 GB/s for the 256GB device and 57.8 GB/s for the 512GB device). CXL's higher thread requirement stems from its higher latency requiring more concurrent requests to hide memory access delays. The 512GB CXL device shows better scaling efficiency than the 256GB device (12.3\% vs 7.5\% at 172 threads relative to 4-thread baseline), suggesting that larger capacity CXL modules benefit more from thread-level parallelism. This observation guides thread pool sizing: applications should provision at least 172 threads for CXL memory to achieve peak bandwidth, nearly 3× more than required for DDR5.

\subsection{Sequential vs Random Access Patterns}

% \begin{figure}
%     \centering
%     \includegraphics[width=\columnwidth]{img/random_vs_seq_comparison.pdf}
%     \caption{Bandwidth comparison between random and sequential access patterns on CXL Node 3 (512GB) across different read-write ratios}
%     \label{fig:random_vs_seq}
% \end{figure}

To understand how access patterns interact with CXL's duplex architecture, we compare random and sequential access performance on the 512GB CXL device (Node 3) using 172 threads and 32GB buffers, as shown in Figure~\ref{fig:microbenchmark}.

\textbf{Observation 5: Random access patterns better utilize CXL's duplex architecture than sequential patterns.} Random access on the 512GB CXL device peaks at 65\% read ratio (62.9 GB/s), maintaining 92\% of peak bandwidth at balanced 50/50 read-write mix. Sequential access peaks at 95\% reads (197.0 GB/s) but drops to only 37\% of peak at 50/50 mix. The write/read performance ratio reveals why: random maintains 0.74× (writes achieve 74\% of reads) while sequential drops to 0.32× due to prefetching and row buffer optimizations that disproportionately benefit sequential reads. This makes random workloads ideal candidates for duplex scheduling as neither reads nor writes can independently saturate bandwidth, creating opportunities for concurrent channel utilization.

\textbf{Observation 6: Access pattern impacts read performance 3.8× more than write performance.} Sequential patterns achieve 186.6 GB/s for reads versus 48.8 GB/s for random reads (282\% improvement), but writes only improve from 36.2 to 59.0 GB/s (63\% improvement). This asymmetry indicates that CXL's write channel operates near practical limits regardless of access pattern, constrained by protocol overhead rather than DRAM row buffer efficiency. For system design, this suggests write-heavy workloads gain minimal benefit from access pattern optimization, while read-heavy workloads should prioritize sequential access where possible.

% \subsection{NUMA Interleaving Performance}

% \begin{figure}
%     \centering
%     \includegraphics[width=\columnwidth]{img/numa_interleave_01_vs_23_fixed_64gb_buffer.pdf}
%     \caption{NUMA interleaving performance comparison between DDR5 (nodes 0-1) and CXL (nodes 2-3) with fixed 64GB buffer across varying thread counts}
%     \label{fig:numa_interleave_buffer}
% \end{figure}

% \begin{figure}
%     \centering
%     \includegraphics[width=\columnwidth]{img/numa_interleave_01_vs_23_fixed_172_threads.pdf}
%     \caption{NUMA interleaving performance comparison between DDR5 (nodes 0-1) and CXL (nodes 2-3) with fixed 172 threads across varying buffer sizes}
%     \label{fig:numa_interleave_threads}
% \end{figure}

% To understand how memory interleaving affects duplex performance, we compare NUMA interleaving between DDR5 nodes (0-1) and CXL nodes (2-3). Interleaving distributes memory accesses across multiple nodes to aggregate bandwidth, but its effectiveness depends on the underlying memory technology characteristics.

\section{Design}
\label{sec:design}

We design \sys as a modular scheduling framework that exploits CXL's duplex capabilities to improve memory bandwidth utilization. This section presents our design rationale, architectural overview, and the key components that enable duplex-aware scheduling.

\subsection{Design Challenges and Rationale}

The need for software-level scheduling in CXL memory systems parallels TCP's bidirectional flow control, where intelligent request reordering maximizes bandwidth utilization. However, unlike network packet scheduling in software, memory accesses are generated directly by hardware, presenting unique optimization challenges. Alternative approaches face fundamental limitations: direct instruction instrumentation incurs prohibitive overhead, compiler-based reordering cannot coordinate across dozens of concurrent CPUs, and hardware solutions lack flexibility for diverse workloads. 

Process-level scheduling provides an effective granularity for exploiting CXL duplexing. The key insight is that by co-scheduling read-intensive and write-intensive processes onto the same CPU cores or NUMA nodes, their naturally interleaved memory requests create a balanced bidirectional traffic pattern at the memory controller. This co-location strategy transforms the traditionally unidirectional memory access patterns of individual processes into a balanced bidirectional stream that maximizes CXL utilization. The approach leverages existing OS infrastructure with minimal overhead while aligning with resource management boundaries for integration with monitoring and control tools. Importantly, our optimization for CXL duplexing is orthogonal to traditional NUMA placement: \sys complements NUMA-aware schedulers by fine-tuning CPU selection within a NUMA node to achieve balanced memory traffic, after the higher-level NUMA placement decision has been made.

\subsection{Architectural Overview}

\begin{figure}
    \centering
\includegraphics[width=\columnwidth]{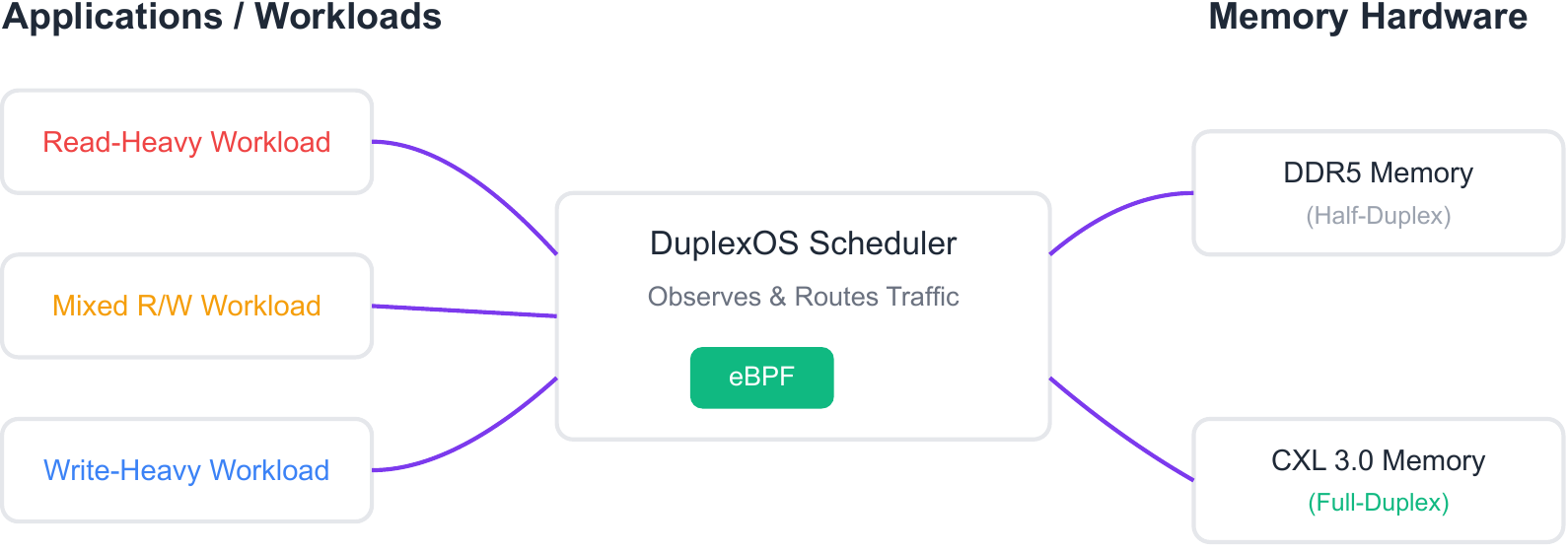}
    \caption{The modular architecture of \sys, showing the interaction between the Observability Infrastructure, Pluggable Policy Engine, and System Integration Interfaces.}
    \label{fig:scheduler}
\end{figure}

\sys comprises three cooperative layers, as illustrated in Figure~\ref{fig:scheduler}: (1) an observability infrastructure for monitoring system state and profiling memory access patterns, (2) a pluggable policy engine that enables runtime selection of scheduling algorithms, and (3) integration interfaces for interacting with standard kernel components including cgroups and memory controllers. This layered design allows \sys to adapt to diverse workloads while facilitating development and deployment of new scheduling strategies.

To ensure broad applicability, our framework is designed as an optimization layer that can be integrated with various underlying scheduling algorithms. It augments the decisions of customized schedulers by providing CXL-duplex-aware hints and overrides. This allows \sys to leverage the strengths of established schedulers—such as topology awareness or real-time guarantees—while layering on our specialized memory bandwidth optimization.

\begin{algorithm}[H]
    \caption{Time Series-Based Scheduler with Oversubscription Detection}
    \begin{algorithmic}[1]
    \STATE \textbf{Initialize:} $TimeSeries \leftarrow \emptyset$, $BandwidthScheduler \leftarrow \emptyset$
    \STATE \textbf{Initialize:} $TaskQueue \leftarrow \emptyset$, $mvruntime \leftarrow 0$
    
    \WHILE{scheduler is running}
        \STATE \textbf{// Phase 1: Update Time Series}
        \STATE $sample \leftarrow CollectSystemMetrics()$
        \STATE $W_t \leftarrow UpdateSlidingWindow(W_t, sample)$
        \STATE $trends \leftarrow CalculateTrends(W_t)$
        
        \STATE \textbf{// Phase 2: Detect Oversubscription}
        \STATE $oversub \leftarrow DetectOversubscription(W_t)$
        \STATE $hint \leftarrow GenerateSchedulingHint(oversub, trends)$
        
        \STATE \textbf{// Phase 3: Dequeue Tasks}
        \WHILE{$task \leftarrow DequeueTask()$}
            \STATE $vruntime \leftarrow UpdateVruntime(task, mvruntime)$
            \STATE $deadline \leftarrow CalculateDeadline(task, vruntime)$
            \STATE $TaskQueue.Insert(task, deadline)$
        \ENDWHILE
        
        \STATE \textbf{// Phase 4: Dispatch Tasks}
        \WHILE{$TaskQueue \neq \emptyset$}
            \STATE $task \leftarrow TaskQueue.PopFirst()$
            \STATE $slice \leftarrow CalculateTimeSlice(hint, task)$
            \STATE $cpu \leftarrow SelectCPU(task, hint)$
            \STATE $DispatchTask(task, cpu, slice)$
        \ENDWHILE
        
        \STATE $NotifyCompletion(|TaskQueue|)$
    \ENDWHILE
    \end{algorithmic}
    \end{algorithm}
\subsection{Observability Infrastructure}

The observability infrastructure enables precise attribution of CXL memory bandwidth to arbitrary profiling scopes, from system-wide metrics down to individual code segments. Unlike traditional PMU-based tools that are limited to CPU-specific events and rigid task or core-level profiling, it provides flexible scope definition and accurate event attribution across heterogeneous execution contexts.

\subsubsection{Flexible Profiling Scopes}

We introduce the CXL Analysis Context (CAX) as a first-class abstraction that enables precise mapping of memory bandwidth measurements to diverse execution scopes. The CAX supports hierarchical scope definitions including system-wide aggregation for global bandwidth analysis for memory topology optimization, process and thread-level attribution for workload characterization, virtual machine boundaries for cloud environments, and code-level scopes for function and loop-level analysis.

The code-level profiling capability is particularly significant for CXL optimization. By leveraging eBPF uprobes to instrument specific functions without kernel modifications or recompilation, the system captures CXL memory bandwidth metrics at function boundaries. For example, a developer can precisely measure the bandwidth characteristics of a matrix multiplication kernel by attaching uprobes to its entry and exit points, obtaining accurate measurements of CXL read bandwidth, write bandwidth, and the read/write ratio during execution. This fine-grained attribution enables targeted optimization of memory-intensive code segments that would be impossible with traditional system-wide or process-level profiling.

\subsubsection{PMU Event Mapping and Attribution}

The infrastructure leverages eBPF to bridge the gap between hardware PMU events and software execution contexts through precise event attribution. Two complementary eBPF mechanisms provide different granularities of analysis: uprobes for function-level profiling that trigger at function boundaries and read PMU counters to measure CXL bandwidth consumption, and programs attached to scheduler events (sched\_switch, sched\_wakeup) for thread-level analysis that capture per-thread bandwidth metrics at context switch boundaries. This dual approach captures true bandwidth consumption at the memory interface while avoiding the inaccuracies inherent in LLC-miss-based bandwidth estimation.

The attribution mechanism maintains a shadow profiling stack that tracks active CAX contexts across all execution levels. When an eBPF hook point is reached, our programs read the relevant PMU counters and attribute the bandwidth delta to the appropriate CAX context. For scheduler events, we capture per-thread bandwidth by reading PMU counters at task switch points, providing fine-grained per-thread analysis without instrumentation. This multi-level attribution enables analysis capabilities, such as understanding how bandwidth consumption of a specific function varies when called from different execution paths or how thread-level bandwidth changes over time.

The observability infrastructure maintains minimal overhead through careful eBPF implementation. PMU reads and bandwidth accounting occur directly in the kernel using lightweight operations, with per-CPU variables avoiding cache contention. For production environments, adaptive sampling automatically adjusts measurement frequency based on system load: reducing overhead during stable patterns and increasing resolution when detecting changes or approaching bandwidth limits.

\subsection{Pluggable Policy Engine}

The \sys policy engine provides a pluggable architecture that allows multiple scheduling algorithms to be implemented, evaluated, and deployed without modifying the core framework. Each policy implements a common interface that receives system state from the observability layer and outputs scheduling decisions to the memory controller. This design enables rapid experimentation with new algorithms while maintaining production stability through policy isolation and runtime switching capabilities.

The framework defines a standard policy interface with three key methods: \texttt{init()} for policy initialization and parameter configuration, \texttt{schedule(state)} for making scheduling decisions based on current system state, and \texttt{update(feedback)} for learning from past decisions. The state object encapsulates queue depths, bandwidth measurements, latency statistics, and cgroup hints, providing a consistent view across all policies. This abstraction allows policies to be developed independently and composed for complex scheduling strategies.

The framework supports dynamic policy switching based on workload changes or administrative commands. Policy transitions are handled through state migration, where the outgoing policy exports its state and the incoming policy imports relevant information. This mechanism enables adaptive scheduling strategies that switch between policies based on observed performance metrics or phase changes in application behavior.

\subsubsection{Scheduling Policies}

We implement a time-series prediction for production workloads, which analyzes historical memory access patterns to proactively optimize task placement. This policy detects workload phase changes and oversubscription conditions through sliding window analysis, enabling preemptive adjustments before performance degradation occurs.

The policy maintains a sliding window of recent system metrics including running threads, CPU utilization, and memory bandwidth measurements. Oversubscription is detected when the average number of running threads exceeds 1.5 per core while CPU utilization remains above 85\%, indicating contention for compute resources. This detection triggers more aggressive scheduling interventions to balance the system load.

\textbf{Scheduling Algorithm:} The scheduling logic is detailed in Algorithm 1, which provides real-time oversubscription detection using time series analysis, adaptive time slicing based on workload volatility, and predictive scheduling using exponentially weighted moving average (EWMA) trend analysis. The algorithm achieves $O(n \log n + m)$ complexity per scheduling iteration for oversubscription detection and task placement.

\subsection{Cgroups Integration and Extensibility}

The \sys framework is designed to integrate cleanly with existing Linux kernel subsystems while providing extensibility for future enhancements. To guide scheduling decisions, \sys relies on explicit hints from applications or system administrators. We use the cgroup interface for this mechanism due to its standardization, security, and seamless integration with container runtimes.

While direct eBPF map interfaces could provide application hints to the scheduler, we adopt cgroups for several reasons. First, purely observability-based tuning suffers from fundamental limitations: it reacts slowly to workload changes, produces imprecise measurements for bursty patterns, and cannot effectively characterize short-lived processes before they complete. These limitations necessitate explicit hints from applications or compilers about expected memory access patterns.

Cgroups provide a suitable mechanism for conveying these hints. They offer a standardized interface that integrates with existing container runtimes (Docker, Kubernetes) and resource management tools, reducing the need for application modifications. The hierarchical nature of cgroups enables natural composition of hints, where system-wide defaults can be overridden at the container level and further refined for specific processes. This design adheres to the principle of least privilege by using a standard, secure kernel interface, preventing unprivileged applications from manipulating system-wide scheduling behavior and ensuring safe deployment in multi-tenant environments.

\section{Implementation}
\label{sec:implementation}

We implement \sys as an eBPF-based scheduler integrated with the Linux 6.14+ sched\_ext framework, intercepting memory allocation and scheduling decisions at multiple system levels. Our implementation specifically targets CXL memory on NUMA node 3 (512GB capacity) to isolate duplex scheduling benefits without the complexity of NUMA placement optimization. This focused approach demonstrates that duplex scheduling can be achieved with acceptable overhead while ensuring full compatibility with existing applications and kernel stability.

\subsection{Observability Implementation}

The observability infrastructure implementation combines eBPF programs and userspace tools. eBPF uprobes are deployed dynamically using the BPF syscall interface, with uprobe programs reading PMU counters through the bpf\_perf\_event \_read\_value helper at function entry and exit points. Bandwidth deltas are calculated and stored in BPF hash maps indexed by CAX context ID.

Thread-level monitoring attaches eBPF programs to the sched\_switch tracepoint using BPF\_PROG\_TYPE\_TRACEPOINT. These programs capture per-thread PMU values at context switch boundaries, maintaining per-CPU ring buffers to minimize overhead. The captured data is aggregated in userspace through memory-mapped BPF maps, providing zero-copy access to bandwidth metrics.

The CAX context management is implemented as a BPF array map with hierarchical indexing. Each CAX entry contains the context type (system, process, thread, function), parent context ID for hierarchy traversal, accumulated read/write bandwidth counters, and timestamp of last update. The hierarchy enables efficient attribution of bandwidth to multiple contexts simultaneously without traversing linked lists in the kernel.

\subsection{Core Scheduler Architecture}

The core scheduler consists of cooperating eBPF programs attached to strategic kernel points, each handling specific aspects of the scheduling lifecycle. The modular design enables independent optimization and testing of components while preserving overall system coherence.

The duplex\_select\_cpu program implements the core mechanism for exploiting CXL duplexing. Triggered before task dispatch, it operates independently of the underlying base scheduler—whether vruntime-based, EDF, or using another algorithm. It examines the task's memory access profile (read/write ratio) from BPF maps and selects a CPU that helps create balanced aggregate traffic. Specifically, if the current CPU cluster is running predominantly read-intensive tasks, the scheduler preferentially places write-intensive tasks on the same cluster to balance the memory controller's request stream. This co-location strategy encourages read and write requests from different processes to arrive at the CXL memory controller in an interleaved fashion, enabling better utilization of both channels. The selection algorithm incorporates hysteresis to reduce excessive migration that could impact CPU cache performance while maintaining the balanced traffic pattern needed for duplex optimization.  We have successfully integrated \sys with over 20 different \texttt{sched\_ext} scheduler implementations, ranging from production-ready schedulers like \texttt{scx\_bpfland}, \texttt{scx\_lavd}, and \texttt{scx\_rusty} to specialized designs such as \texttt{scx\_nest} for frequency optimization and \texttt{scx\_layered} for multi-layer scheduling policies, demonstrating the framework's versatility across diverse scheduling paradigms.

The duplex\_enqueue program updates queue statistics and triggers rebalancing when thresholds are exceeded, using separate per-CPU data structures to minimize cache contention. Each CPU tracks local read and write queue depths that are periodically synchronized to global counters using atomic operations. The program implements adaptive batching that groups multiple requests to amortize synchronization overhead while preserving reasonable latency.

\subsection{Policy Engine Implementation}

The policy engine is implemented as a set of loadable eBPF programs that can be dynamically switched at runtime. Each policy is compiled as a separate BPF object file containing the scheduling logic. Our framework can enhance multiple base schedulers, and our production deployment uses the time-series predictive policy (Algorithm 1) which provides the best balance of performance and adaptability when combined with the underlying scheduler. The framework loads the appropriate policy based on configuration and workload characteristics, with seamless handoff between policies through shared state structures. The time-series policy's sliding window analysis and oversubscription detection are implemented using BPF ring buffers and per-CPU variables to maintain the window of samples with low overhead.

\section{Evaluation}
\label{sec:evaluation}

We evaluate \sys using both synthetic microbenchmarks and real-world applications to demonstrate duplex-aware scheduling effectiveness across diverse workload characteristics. Our evaluation addresses two key research questions:

\begin{itemize}
    \item{RQ1: How much can duplex-aware scheduling improve CXL memory performance over baseline CXL?} We quantify the performance gains achievable through intelligent scheduling policies compared to default Linux scheduler.
    \item{RQ2: What are the performance impacts across diverse real-world applications?} We evaluate \sys on production workloads including key-value stores, large language model inference, and vector databases to understand both benefits and trade-offs.
\end{itemize}

\subsection{Experimental Setup}

Experiments use the hardware platform from Section~\ref{sec:characterization} with Ubuntu 25.04 and Linux kernel 6.14.0-27-generic including \sys eBPF scheduler extensions. Our evaluation methodology was designed to identify the optimal combination of baseline scheduling algorithm and our \sys optimization policies. We integrated \sys as an enhancement layer on top of several production-ready \texttt{sched\_ext} schedulers, evaluating multiple combinations to determine the configuration that provides the most robust performance across our target data-intensive workloads. The baseline for all comparisons is the EEVDF default Linux kernel scheduler without any CXL-specific optimizations. All evaluations employ the best-performing configuration identified through our preliminary experiments: \sys with the time-series predictive policy. All workloads run on NUMA node 3's 512GB CXL memory to isolate duplex scheduling benefits. We employ 172 concurrent threads, NUMA-aware memory binding, and disabled CPU frequency scaling, with results averaged over 3 runs.

\subsection{Microbenchmark Results}

\begin{figure}
    \centering
\includegraphics[width=\columnwidth]{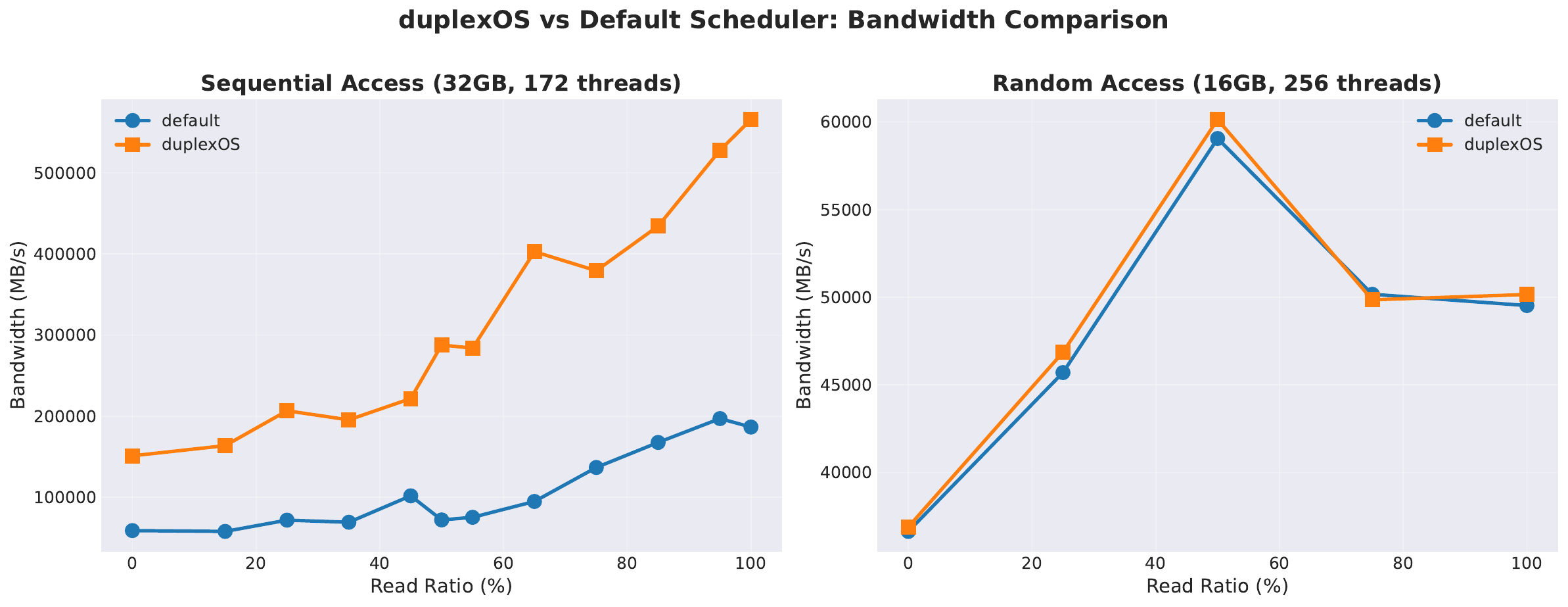}
    \caption{Microbenchmark performance comparison between \sys and Linux CFS (baseline) across sequential and random access patterns. Y-axis shows bandwidth in GB/s (1 GB = $10^9$ bytes). Tests performed on CXL memory (NUMA node 3) using --membind=3.}
    \label{fig:microbenchmark}
\end{figure}

The microbenchmark evaluation systematically explores the parameter space of memory access patterns to understand scheduler behavior under controlled conditions. Custom benchmarks generate precise access patterns while measuring achieved bandwidth, latency distributions, and resource utilization with minimal measurement overhead.

Sequential access patterns with 32GB working set and 172 threads demonstrate significant benefits of duplex scheduling. The time-series predictive policy detects phase transitions between read and write operations through its sliding window analysis (Algorithm 1), enabling proactive task migration before queue imbalances occur. \sys achieves up to 195.9\% improvement over the Linux CFS baseline for specific read-write ratios, with an average improvement of 95.8\% across all ratios tested. The superior performance of sequential patterns compared to random access stems from the predictability that enables the time-series model to accurately forecast future access patterns using EWMA trend analysis. Random access patterns with 16GB working set and 256 threads present a more challenging scenario, where \sys shows modest improvements of 1.2\% on average with a maximum improvement of 1.8\%. Here, the time-series policy's oversubscription detection (Equation 4) prevents performance degradation by maintaining stable scheduling decisions despite unpredictable access patterns.

Across both sequential and random workloads, \sys deliver an average improvement of 48.5\% across all tested configurations, as shown in Figure~\ref{fig:microbenchmark}. These results validate that duplex-aware scheduling can nearly double memory system performance for suitable workloads while maintaining stability for challenging access patterns.

\subsection{Redis Performance Analysis}

\begin{figure*}
    
    \centering
\includegraphics[width=\textwidth]{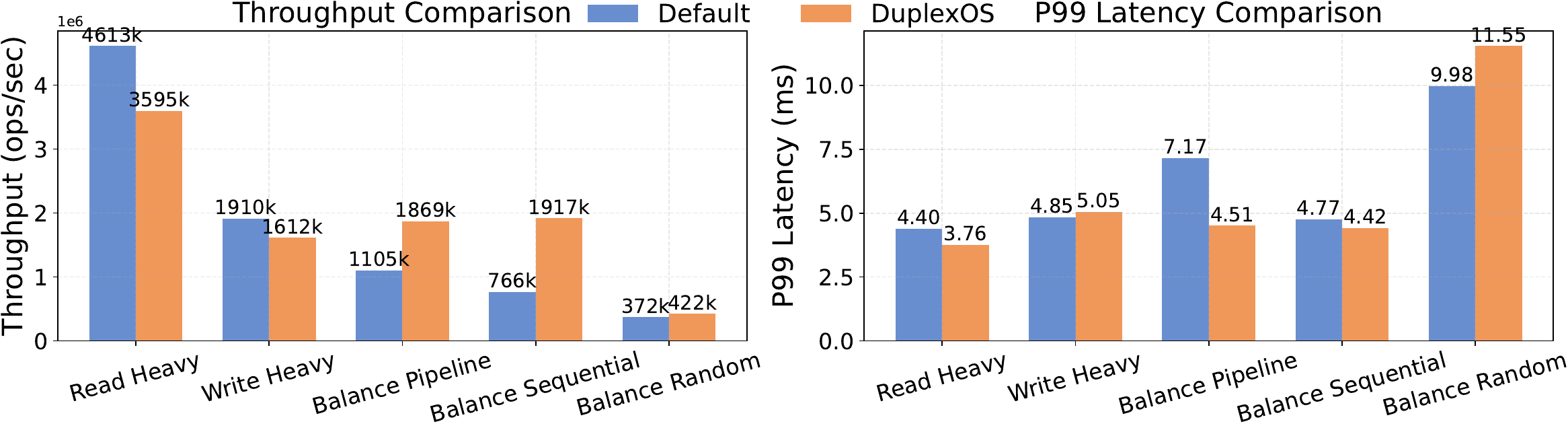}
    \caption{Redis performance comparison between Linux CFS (baseline) and \sys schedulers on CXL memory (NUMA node 3, 512GB) using --membind=3. Error bars show 95\% confidence intervals over 10 runs.}
    \label{fig:redis}
\end{figure*}
Redis represents a key workload for evaluating duplex scheduling benefits in production scenarios. We benchmark Redis 7.0 with multi-threaded I/O using memtier\_benchmark, testing 320 concurrent clients (32 threads × 10 clients) across five distinct access patterns on a 256GB dataset deployed on CXL memory.

\subsubsection{Throughput Analysis}
The throughput results in Figure~\ref{fig:redis} reveal workload-dependent performance enabled by the time-series policy's adaptive behavior. Read-heavy workloads (1:10 SET:GET ratio) show a 22\% throughput reduction under \sys, dropping from 4.61M to 3.60M ops/sec, while write-heavy patterns (10:1 ratio) decrease by 16\% from 1.91M to 1.61M ops/sec. The time-series policy detects these unidirectional patterns through its sliding window analysis and automatically reduces scheduling intervention to minimize overhead. However, balanced workloads demonstrate substantial gains: pipelined operations (16 commands batched) improve by 69\% from 1.11M to 1.87M ops/sec, as the predictive model anticipates the alternating read-write pattern of pipelined commands. Sequential access patterns achieve 150\% improvement from 0.77M to 1.92M ops/sec, where the time-series forecasting accurately predicts phase transitions. Gaussian-distributed random access shows modest 14\% gains from 0.37M to 0.42M ops/sec. Overall, \sys delivers 7.4\% average throughput improvement across all patterns.

\subsubsection{Latency Characteristics}
P99 latency analysis presents complementary insights. Read-heavy workloads achieve 15\% latency reduction (4.40ms to 3.76ms) despite lower throughput, indicating improved consistency. Write-heavy patterns show minimal degradation (4\% increase to 5.05ms). Pipelined operations benefit most with 37\% latency improvement (7.17ms to 4.51ms), while sequential patterns reduce by 7\% (4.77ms to 4.42ms). Random Gaussian workloads experience 16\% latency increase (9.98ms to 11.55ms), suggesting scheduling overhead for unpredictable access patterns. The average P99 latency improves by 6\%, demonstrating that duplex scheduling enhances predictability alongside throughput.

Sequential and pipelined patterns achieve the best results (150\% and 69\% throughput gains) due to balanced traffic that maximizes concurrent channel utilization. The 22\% regression for read-heavy workloads confirms duplex scheduling overhead for unidirectional traffic. The cgroup hint mechanism enables selective duplex scheduling based on workload characteristics.

\subsection{Large Language Model Inference Performance Analysis}

\begin{figure}
    \centering
\includegraphics[width=\columnwidth]{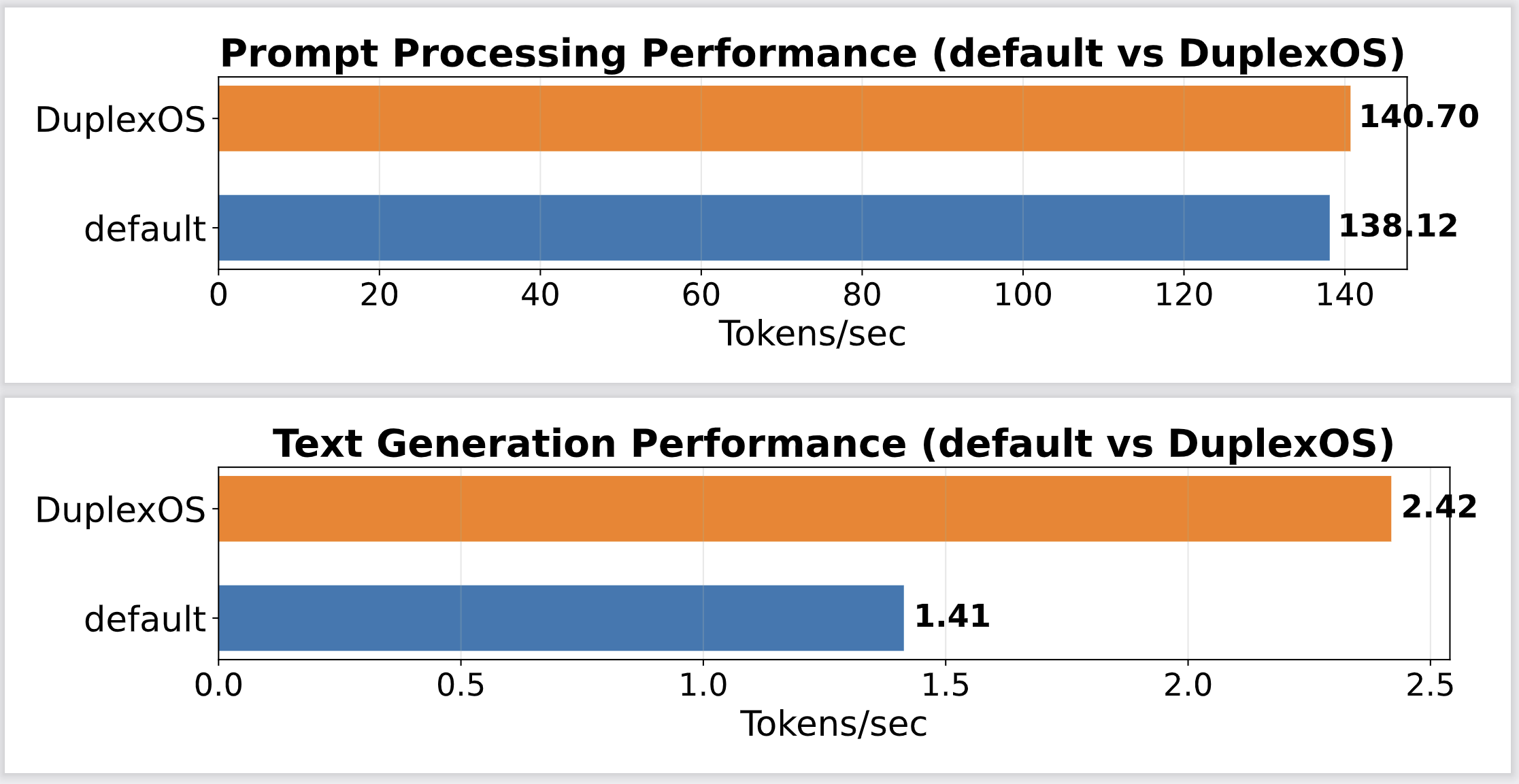}
    \caption{LLAMA inference performance with \sys on CXL memory (NUMA node 3, 512GB) using --membind=3. Throughput measured in tokens/second.}
    \label{fig:llama}
\end{figure}

% \begin{figure}
%     \centering
% \includegraphics[width=\columnwidth]{img/roofline.jpg}
%     \caption{\sys LLAMA Roofline Result}
%     \label{fig:roofline}
% \end{figure}

We evaluate LLM inference using LLAMA3 70B and Deepseek V3-0324 671B models with weights and activations in CXL memory.

The model's memory access patterns vary across transformer layers, creating opportunities for scheduling. Attention computations generate approximately 85\% read traffic as they access cached key-value pairs from previous tokens in the sequence. Feed-forward network layers exhibit more balanced patterns with 60\% reads and 40\% writes as they transform representations. The total memory footprint includes 660GB for Intel AMX INT8 model parameters plus dynamic activation storage that scales with batch size and sequence length.

For prompt processing, \sys achieves 140.70 tokens/second compared to 138.12 tokens/second with Linux CFS, a modest 1.8\% improvement due to the read-heavy nature of initial context loading. The time-series policy's oversubscription detection (Equation 4) identifies the compute-bound nature of this phase and minimizes scheduling intervention. However, text generation shows substantial gains: \sys delivers 2.42 tokens/second versus 1.41 tokens/second for Linux CFS, representing a 71.6\% improvement. The time-series model excels here by predicting the regular alternation between attention layers (85\% read) and feed-forward layers (60\% read, 40\% write), proactively co-scheduling tasks to maintain balanced bidirectional traffic throughout the transformer pipeline.

\subsection{Vector Database Performance Analysis}

\begin{figure}
    \centering
\includegraphics[width=\columnwidth]{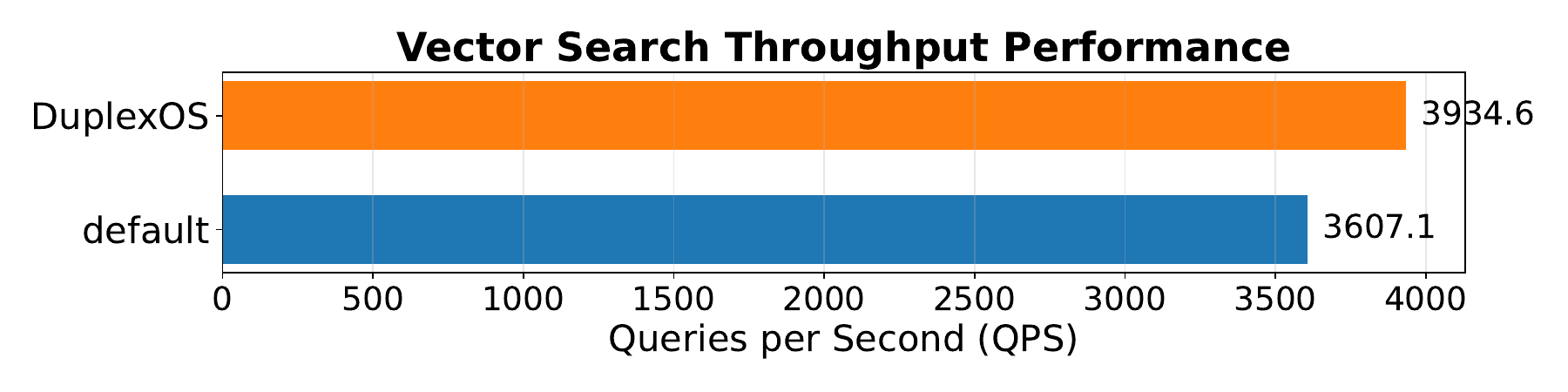}
    \caption{PyVSAG vector database performance comparison between Linux CFS (baseline) and \sys schedulers on CXL memory (NUMA node 3, 512GB) using --membind=3. Error bars show 95\% confidence intervals.}
    \label{fig:vectordb}
\end{figure}

Vector databases are essential for modern AI applications including similarity search, recommendation systems, and retrieval-augmented generation. We evaluate PyVSAG, a high-performance vector search library, using HNSW (Hierarchical Navigable Small World) indexing on CXL memory. The benchmark tests k-nearest neighbor search performance with 50,000 128-dimensional vectors and 1,000 query operations.

Figure~\ref{fig:vectordb} presents the performance comparison between Linux CFS and \sys schedulers. \sys achieves 3,935 queries per second (QPS) compared to 3,607 QPS for Linux CFS, representing a 9.1\% throughput improvement. Average query latency decreases from 276.4$\mu$s to 253.4$\mu$s, an 8.3\% reduction. These improvements stem from better memory access scheduling during the graph traversal operations inherent in HNSW search.

The HNSW algorithm exhibits mixed read-write patterns during search operations: reads dominate when traversing the hierarchical graph structure to locate nearest neighbors, while writes occur during distance computation caching and result aggregation. This access pattern aligns well with CXL's duplex architecture. The 9\% performance gain, while modest compared to sequential workloads, provides meaningful improvements for latency-sensitive vector search applications where microsecond-level reductions directly impact user experience. The consistent improvement across both throughput and latency metrics validates that duplex scheduling benefits extend to complex graph-based data structures beyond simple key-value operations.

\section{Related Work}
\label{sec:related}

The emergence of CXL has catalyzed extensive research across multiple dimensions of memory system design. Early CXL characterization work focused on quantifying performance metrics and demonstrating feasibility~\cite{gouk2022direct,sun2023demystifying,wahlgren2024understanding}. Gouk et al. presented DirectCXL, one of the first systems to demonstrate CXL memory in practice, achieving near-DRAM performance for memory-intensive applications~\cite{gouk2022direct}. However, their evaluation focused on isolated latency and bandwidth measurements without considering the mixed read-write workloads where duplex benefits manifest~\cite{li2023pond,maruf2023tpp}. This work identifies the duplex opportunity overlooked in prior studies. The software stack for CXL systems has evolved to support new usage models and deployment scenarios. Meta's TPP (Transparent Page Placement) pioneered page-level management for CXL memory, using hotness tracking to optimize data placement between local and CXL-attached memory~\cite{maruf2023tpp}. While TPP reduces average access latency through intelligent placement, it operates at page granularity and does not consider request-level scheduling that could exploit duplex channels~\cite{yan2019nimble,agarwal2020thermostat}. This work complements TPP, as duplex scheduling could enhance TPP's effectiveness by improving CXL bandwidth utilization for pages placed in remote memory.

Traditional memory scheduling research provides the foundation for our work, though prior algorithms assume half-duplex operation~\cite{rixner2000memory,mutlu2008parallelism,kim2010thread}. The seminal FR-FCFS algorithm prioritizes requests to open DRAM rows, improving row buffer hit rates but not addressing duplex opportunities~\cite{rixner2000memory}. PAR-BS groups requests into batches to exploit bank-level parallelism while maintaining fairness, but its batching approach actually reduces duplex utilization by serializing operations~\cite{mutlu2008parallelism,mutlu2007stall}. Prior work on application-aware memory management demonstrates the value of incorporating application knowledge into system decisions~\cite{leverich2014reconciling,lo2015heracles,verma2015large}. Heracles achieves predictable latency for latency-critical services through isolation and resource management~\cite{lo2015heracles}. The cgroup-based hint mechanism extends these ideas to memory channel scheduling, providing a familiar interface for application guidance~\cite{menage2007cgroups,heo2013cgroup}.

\section{Conclusion}

We presented \sys, a system that exploits CXL's full-duplex architecture to improve memory system performance. The characterization revealed that CXL achieves 55-61\% bandwidth improvement at balanced read-write ratios compared to DDR5's relatively flat performance across different ratios, identifying a software gap limiting this potential. We developed a scheduling framework with multiple policies including a cgroup-based hint mechanism for application-aware optimization, implemented as a Linux kernel extension using eBPF. Evaluation demonstrates 7.4\% average improvement in Redis throughput, 71.6\% improvement for LLM text generation, and 9.1\% for vector databases, with specific workloads like sequential Redis operations achieving up to 150\% improvement. These results suggest that architectural innovations benefit from software co-design, and the techniques developed may be applicable to future memory technologies with similar characteristics.

\bibliographystyle{ACM-Reference-Format}
\bibliography{cite}

\end{document}